\documentclass[lettersize,journal]{IEEEtran}

\usepackage{amsmath,amssymb,amsfonts,bm}
\usepackage{array,booktabs,multirow,makecell,longtable,tabularx,adjustbox,ragged2e}
\setcellgapes{3pt}

\usepackage{graphicx}
\usepackage[caption=false,font=normalsize,labelfont=sf,textfont=sf]{subfig}
\usepackage{stfloats}

\usepackage{algorithm,algorithmic}

\usepackage[justification=centering]{caption}
\usepackage[utf8]{inputenc}

\usepackage{cite}
\usepackage{tabularx} 
\usepackage{xcolor}
\usepackage{url}
\usepackage{verbatim}
\usepackage{ragged2e}
\usepackage{changepage}
\usepackage{pifont}

\usepackage[colorlinks=true,linkcolor=blue,citecolor=blue,urlcolor=blue]{hyperref}

\begin{document}

\title{Aerial Agentic AI: Synergizing LLM and SLM for Low-Altitude Wireless Networks}

\author{Li Dong, Feibo Jiang, \textit{Senior Member, IEEE}, Kezhi Wang, \textit{Senior Member, IEEE}, Cunhua Pan, \textit{Senior Member, IEEE}, Dong In Kim, \textit{Life Fellow, IEEE}, Ekram Hossain, \textit{Fellow, IEEE}
}

\maketitle

\begin{abstract}
Low-Altitude Wireless Networks (LAWNs), composed of Unmanned Aerial Vehicles (UAVs) and mobile terminals, are emerging as a critical extension of 6G. However, applying Large Language Models in LAWNs faces three major challenges: 1) Computational and energy constraints; 2) Communication and bandwidth limitations; 3) Real-time and reliability conflicts. To address these challenges, we propose Aerial Agentic AI, a hierarchical framework integrating UAV-side fast-thinking Small Language Model (SLMs) with BS-side slow-thinking Large Language Model (LLMs). First, we design SLM-based Agents capable of on-board perception, short-term memory enhancement, and real-time decision-making on the UAVs. Second, we implement a LLM-based Agent system that leverages long-term memory, global knowledge, and tool orchestration at the Base Station (BS) to perform deep reasoning, knowledge updates, and strategy optimization. Third, we establish an efficient hierarchical coordination mechanism, enabling UAVs to execute high-frequency tasks locally while synchronizing with the BS only when necessary. Experimental results validate the effectiveness of the proposed Aerial Agentic AI.
\end{abstract}


\section{Introduction}
With the rapid development of the low-altitude economy, Low-Altitude Wireless Networks (LAWNs) composed of Unmanned Aerial Vehicles (UAVs), aerial Base Stations (BSs), and mobile terminals are emerging as an important extension of 6G \cite{10274880}. During missions such as inspection, monitoring, and emergency response, UAVs continuously collect and generate massive volumes of multimodal sensing data, including aerial images and videos, radar and inertial navigation information, environmental and target sensing data, as well as network state information.

In such highly dynamic and uncertain low-altitude scenarios, conventional edge intelligence approaches dominated by discriminative inference struggle to simultaneously satisfy the complex requirements of real-time environment modeling, semantic communications, and intelligent decision-making. To address these limitations, generative edge intelligence has emerged as a promising paradigm. Generative edge intelligence refers to the deployment of generative models at the network edge, enabling intelligent content generation, decision-making, and interaction in close proximity to data sources and end devices \cite{10955732}.

\textcolor{black}{Large Language Models (LLMs) are foundation generative models pre-trained on massive-scale corpora with billions to hundreds of billions of parameters.} In recent years, LLMs have enabled intelligent communications by providing strong semantic understanding and reasoning capabilities\cite{11370176}. However, directly migrating such LLMs to LAWNs still faces a series of fundamental challenges:

\subsubsection{Computational and Energy Constraints}
\textcolor{black}{LLMs typically comprise tens or even hundreds of billions of parameters, and their inference and generation processes demand extremely high computational resources and parallel processing capabilities. However, in LAWNs, aerial nodes are subject to stringent constraints on both computing capacity and energy supply, making it difficult to sustainably support such computation-intensive inference and generation workloads \cite{10258360}. This mismatch between computational demand and energy availability constitutes the primary challenge in migrating LLMs to LAWNs.}

\subsubsection{Communication and Bandwidth Limitations}
\textcolor{black}{Generative tasks often require the transmission of long contexts and multimodal data, along with frequent interactions with external memory and knowledge bases. Nevertheless, LAWNs commonly suffer from limited bandwidth, unstable connectivity, and link congestion, which makes it difficult to sustain continuous uplink and downlink transmission of large-scale data during the generation process \cite{10835069}. This limitation directly increases end-to-end latency and degrades generation quality, and in severe cases may lead to service interruption or forced performance degradation.}

\subsubsection{Real-Time and Reliability Conflicts}
\textcolor{black}{Tasks in LAWNs impose strict requirements on the latency and executability of the “perception–decision–control” closed loop. However, the generation process of LLMs is typically characterized by long inference times. Moreover, under rapidly changing environments, the lack of effective external tool (function) invocation mechanisms may result in incorrect tool selection or unstable inference sequences, leading to output fluctuations, hallucinations, and reduced consistency \cite{10384606}. As a consequence, the reliability of the generated results becomes difficult to guarantee.}

\setcounter{subsubsection}{0}

To address the above challenges, this study proposes Aerial Agentic AI for LAWNs. 
\textcolor{black}{Agentic AI refers to a LLM-based paradigm in which agents operate in a goal-driven closed loop and improve over time via memory and feedback.}
The core idea is to deploy Small Language Model (SLM)-based agents with autonomous perception, reasoning, decision-making, and execution capabilities on aerial edge nodes such as UAVs. By leveraging local inference, short-term memory, and native tool invocation of “fast-thinking” SLMs, the proposed framework maintains an on-board reasoning loop and enables immediate responses at the aerial side. When link conditions permit, these agents further collaborate with “slow-thinking” LLMs deployed at BSs to perform global tool orchestration, deep reasoning, and long-term memory updates. In this way, low-latency, high-reliability, and continuously evolving intelligent service provisioning can be achieved in highly dynamic low-altitude communication environments. Specifically, the proposed framework comprises the following components:

\subsubsection{SLM and LLM Collaboration}
To cope with the stringent computational and energy constraints of UAVs, Aerial Agentic AI adopts a hierarchical reasoning paradigm. SLMs are deployed on UAVs to perform high-frequency perceptual understanding and instantaneous action decision-making with low computational overhead, thereby maintaining a stable closed loop under tight resource budgets. In contrast, LLMs are operated at BSs and assume the role of slow thinking, handling complex reasoning, global planning, and policy optimization. The resulting refined and structured policies are then transmitted to the UAVs for synchronized execution. 
\textcolor{black}{Moreover, we further develops an aerial task-oriented capability transfer scheme to migrate mission-specific knowledge and skills from ground-side LLMs to on-board SLMs.}

\subsubsection{Short-Term and Long-Term Memory Coordination}
To address the limitations of bandwidth and the instability of low-altitude links, Aerial Agentic AI establishes a two-tier memory architecture. On the UAV side, only task-relevant short-term context and episodic experiences are retained to support continuous decision-making under weak or even disconnected network conditions. On the base-station side, long-term memory and global knowledge are maintained to accumulate cross-task experience and update knowledge over time. The most essential information is fed back to UAVs through low-overhead summary or index synchronization. This mechanism transforms air–ground interaction from “large-scale context backhaul” into “semantic summarization and index synchronization,” thereby reducing communication overhead while preserving memory availability and enabling continual experience refinement.

\subsubsection{On-Board Tool Invocation and Off-Line Tool Orchestration}
To enhance the online executability and resource controllability of aerial agents in dynamic tasks, Aerial Agentic AI introduces a hierarchical tool framework. On UAVs, SLMs perform rapid tool selection primarily over on-board native interfaces and lightweight operators (e.g., cameras and radar, sensors, flight control APIs, and lightweight detection and tracking operators), enabling the “perception–decision–control” loop to be completed in a deterministic and low-latency manner. At BSs, high-capacity LLMs orchestrate and schedule complex tool chains across tasks and UAVs (e.g., multi-UAV path planning, cooperative task allocation, and global resource scheduling), and then dispatch structured, executable task instructions to the aerial side for execution. This hierarchical mechanism preserves basic closed-loop capabilities under weak network conditions, while enabling stronger task collaboration and global optimization once connectivity is restored.

The remainder of this paper is organized as follows. Section II introduces Aerial Agentic AI and its applications. Section III reviews SLM advantages and key optimization techniques. Section IV details the proposed framework. Section V presents experiments. Section VI discusses open issues, and Section VII concludes the paper.

\section{Characteristics and Application Scenarios}
\subsection{Characteristics}

Aerial Agentic AI proposes an air–ground cooperative agent architecture with a clear division of labor: UAVs employ fast-thinking systems (SLMs) for rapid decision-making and immediate execution, while BSs run slow-thinking systems (LLMs) for global planning, strategic reflection, and complex tool orchestration. 
Hence, Aerial Agentic AI integrates three key characteristics:

\subsubsection{Autonomy}
The aerial side prioritizes independent task-level closed-loop operation. Even under link degradation or disconnection, UAVs can maintain a basic ``perception–decision–control" loop using SLMs, short-term memory, and on-board tools.

\subsubsection{Collaboration}
Air–ground and multi-UAV collaboration is achieved through low-overhead state sharing. The aerial side transmits critical semantic information and local states, while the ground side performs complex reasoning, long-term memory updates, and global optimization.

\subsubsection{Consistency}
\textcolor{black}{The slow-thinking system defines a rigorously specified constraint boundary for the fast-thinking system, thereby enforcing that the SLM’s statistical inferences comply with the physical laws governing LAWNs and ensuring the physical feasibility and logical self-consistency of long-term planning.} 

\subsection{Application Scenarios}

The Aerial Agentic AI can be categorized into the following six representative application scenarios:

\subsubsection{Emergency Communication Support}
In scenarios such as disaster rescue, network outage recovery, or network congestion, UAVs act as temporary relays or aerial BSs and must rapidly complete takeoff, deployment, coverage planning, and link reconfiguration. On-board Agentic AI enables UAVs to perceive coverage blind spots and link degradation in real time, and to adaptively adjust altitude, flight trajectories, and backhaul strategies to ensure the continuous availability of critical services.

\subsubsection{Urban Monitoring and Data Backhaul}
For traffic monitoring, security surveillance, and crowd event observation, UAVs are required to continuously deliver actionable information streams under fluctuating bandwidth conditions. Aerial Agentic AI performs on-board event cue extraction and key evidence selection (e.g., key frames, critical video segments, and alert tags), and adaptively determines the uplink granularity according to available bandwidth, thereby avoiding congestion caused by full raw-data backhaul.

\subsubsection{Target Tracking and Re-acquisition}
In search-and-rescue operations and vehicle or personnel tracking, UAVs must maintain continuous tracking under highly dynamic conditions and occlusions, and rapidly re-acquire targets upon loss. Aerial Agentic AI maintains short-term contextual awareness of target motion patterns and online generates micro-level tracking plans and search strategies (e.g., expanding search sectors, adjusting altitude and speed, or switching sensors), thereby improving task continuity.

\subsubsection{Fine-Grained Inspection}
For inspection tasks involving power lines, pipelines, and bridges, UAVs must translate inspection objectives into executable actions, such as approaching targets, precise alignment, stable hovering, and supplementary imaging of critical components, while rapidly adapting observation strategies under occlusion, glare, or vibration. Aerial Agentic AI enables preliminary defect cue screening and evidence organization on-board, improving inspection efficiency and reducing backhaul pressure.

\subsubsection{3D Environment Reconstruction}
For applications such as digital twins,  and environmental reconstruction, UAVs are required to collect data from multiple viewpoints and altitudes while ensuring coverage completeness. Aerial Agentic AI can online assess information gaps and sampling value, dynamically adjust sampling trajectories and imaging density, and prioritize the transmission of key frames, critical segments, and structured semantic summaries under poor link conditions, thereby supporting efficient ground-side reconstruction.

\subsubsection{Voice-Based Command and Interaction}
In on-site command, emergency coordination, and operation and maintenance scenarios, voice interaction provides a low-barrier human–machine interface, allowing operators to issue natural-language commands for takeoff and landing, task switching, area designation, and status querying. Aerial Agentic AI enables lightweight on-board speech command understanding and intent parsing, mapping instructions into executable flight control, communication, and payload-control actions.

\section{Aerial SLMs}
\subsection{Advantages of Aerial SLMs}
In LAWNs, aerial SLMs refer to compact language models that can be efficiently deployed on the resource-constrained intelligent units of aerial nodes (e.g., UAVs), enabling low-latency and energy-efficient on-device inference. \textcolor{black}{Although context engineering can improve output quality, relying solely on context-level optimization is insufficient to meet the stringent on-board resource budgets and deterministic real-time requirements of aerial platforms; therefore, the SLM remains a key enabler for controllable low-latency inference \cite{zhao2026wireless}.}
Aerial SLMs exhibit the following key characteristics:

\subsubsection{High Capability Density}
SLMs at the scale of approximately 7B parameters have demonstrated the ability to approach, and in some tasks even reach, the performance of 70B-scale LLMs in capabilities such as code generation, tool invocation, and instruction following.
This reflects a significantly higher efficiency in parameter utilization and capability density. For UAV-side deployment, this implies that practical reasoning and generation capabilities can be achieved with substantially lower computational overhead, thereby supporting on-board autonomous decision-making and interactive tasks.

\subsubsection{Strong Adaptability}
SLMs are easier to deploy locally on UAV platforms, greatly reducing dependence on cloud resources and network conditions and thus better satisfying real-time interaction requirements. Moreover, SLMs are well suited to rapid alignment and convergence on specific tasks or workflows through lightweight fine-tuning, enabling faster scenario-specific deployment and adaptation to the edge environment, which is typically characterized by well-defined tasks and rapid changes.

\subsubsection{Low Inference Cost}
SLMs impose lower demands on computation, memory, and energy, thereby reducing hardware requirements and resource consumption for concurrent inference. Their reduced computational complexity and power consumption significantly decrease the overall cost of inference while improving service capacity per unit cost. Consequently, SLMs offer superior economic efficiency and scalability for deployment in LAWNs.

\subsection{Key Techniques}
\begin{figure}[htbp]
	\centering
	\includegraphics[width=8cm]{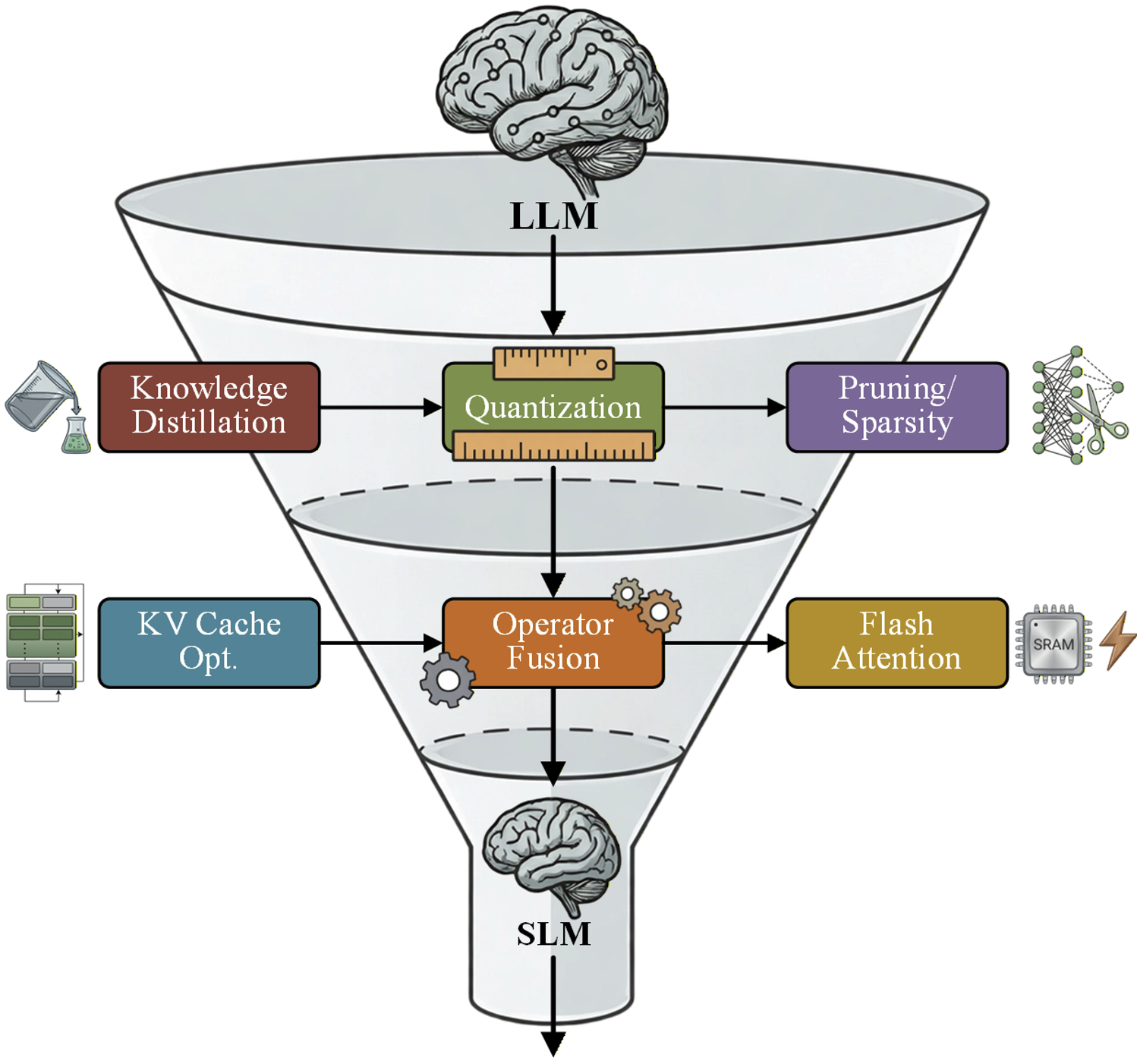}
	\caption{Key Technologies for SLM Optimization.}
	\label{fig:fig21}
\end{figure}

The design of Aerial SLMs requires a systematic set of techniques spanning multiple dimensions, including model architecture and inference mechanisms, to reconstruct the capabilities of LLMs into compact forms tailored to closed-loop flight missions as shown in Fig. \ref{fig:fig21}. 
The key techniques include the following.

\subsubsection{Knowledge Distillation}
On the UAV side, the goal of distillation is to transfer the decision-making capabilities of ground-based LLMs, trained for specific aerial tasks to SLMs. This allows the models to maintain task-usable output quality despite a significant reduction in parameter scale\cite{moslemi2024survey}. Through soft-label distillation and task-specific distillation, SLMs can achieve higher capability density under limited computational resources, thereby improving executability and operational safety.

\subsubsection{Model Quantization}
By compressing model weights and activations from floating-point representations to INT8 or INT4, quantization significantly reduces model size and computational cost. This approach is particularly effective on aerial platforms that provide native support for low-precision arithmetic, yielding more stable energy-efficiency gains. SLMs typically adopt precision-aware hybrid quantization strategies—maintaining higher precision for critical layers while applying low-bit quantization to others—to mitigate the risk of task misjudgment\cite{lin2024awq}.

\subsubsection{Model Compression}
Structured pruning (e.g., channel-wise, attention-head, or layer-wise pruning) directly reduces matrix dimensions and memory access, making it more suitable for efficient execution on embedded GPUs or NPUs \cite{ma2023llm}. Low-rank approximation reduces the number of multiply–accumulate operations by decomposing large matrices into low-rank products, thereby lowering inference cost without severely degrading representational capacity. These techniques are often combined with distillation or quantization to compensate for potential performance loss.

\subsubsection{KV Cache Optimization}
KV caches are a primary source of memory pressure for long-context inference in SLMs and directly affect peak memory usage, tail latency, and the risk of out-of-memory failures \cite{li2024snapkv}. Techniques such as KV cache quantization to reduce footprint, sliding-window mechanisms to limit context growth, and importance-based token selection to retain only critical cached entries can effectively reduce redundancy. 

\subsubsection{Operator Fusion}
Operator Fusion (OF) merges adjacent operators to reduce intermediate tensor reads/writes and kernel launch overhead \cite{salmani2025llm}. On embedded GPU or NPU platforms, this can substantially increase throughput and reduce energy consumption. For SLMs, OF is particularly beneficial for latency reduction, as it minimizes frequent memory transfers and scheduling jitter, leading to more deterministic inference times that better satisfy the stringent latency requirements of aerial tasks.

\subsubsection{Flash Attention}
Flash Attention (FA) improves end-to-end inference efficiency by computing attention in blocks and fusing key operations such as softmax and matrix multiplication, thereby significantly reducing memory access and bandwidth consumption for intermediate attention matrices \cite{dao2022flashattention}. In long-context and real-time decoding scenarios typical of SLMs, this technique effectively alleviates memory bottlenecks in attention operators, reduces inference latency and its variance, and enhances attention efficiency. 

In summary, optimization techniques for SLMs exhibit multi-dimensional and multi-layered characteristics, encompassing both model-level methods (distillation, quantization, and compression) and inference-level optimizations (cache management, OF, and FA). 

\section{Aerial Agentic AI Framework}
The design objective of Aerial Agentic AI is to preserve the autonomy, adaptability, and real-time responsiveness of aerial agents under stringent constraints on computation, storage, and communication. The overall framework follows three core principles. First, the key capabilities of ground-side LLMs are transferred to SLMs in the form of usable capability units. Second, core components—such as tool invocation, memory, reflection, knowledge management, and reasoning and planning—are implemented in a lightweight and modular manner to satisfy the deterministic latency and resource budgets of aerial nodes. Third, air–ground collaboration is leveraged to achieve complementary capabilities and continuous enhancement, forming a system paradigm in which the aerial side maintains a usable closed loop, while the ground side performs end-to-end global optimization.

\subsection{Capability Transfer from LLMs to SLMs}
\begin{figure*}[htbp]
	\centering
	\includegraphics[width=16cm]{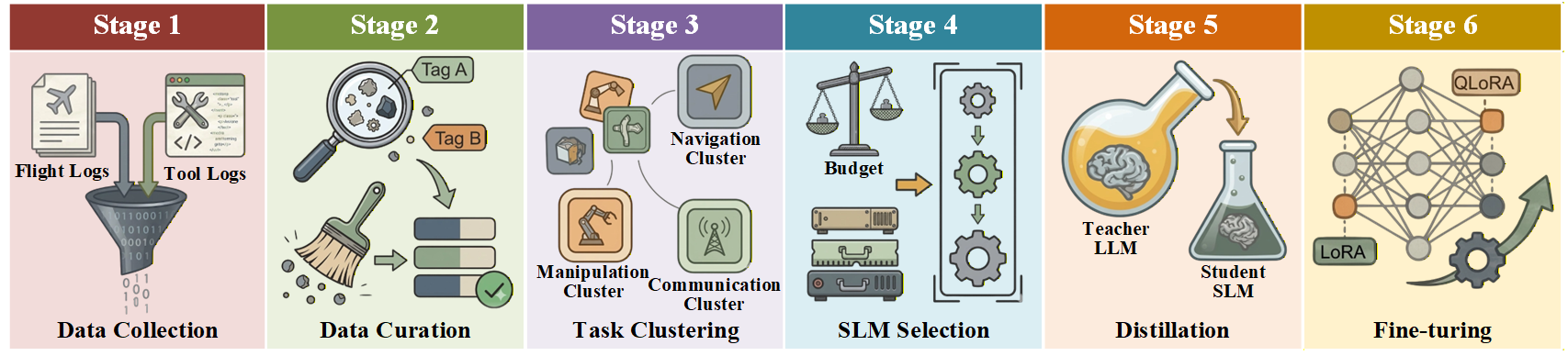}
	\caption{Capability Transfer from LLMs to SLMs.}
	\label{fig:fig31}
\end{figure*}

The process of transferring capabilities from LLMs to SLMs is outlined as follows and is shown in Fig. \ref{fig:fig31}.

\textit{S1. Data Collection: }
On both the UAV and BS sides, reproducible behavioral trajectories of agents during real missions are recorded, including input prompts, output responses, sequences of tool or interface invocations (e.g., camera, radar, flight control, and CSI), and key context summaries. These data are used to construct task datasets for subsequent modeling and training. 

\textit{S2. Data Curation: }
Based on the data streams collected in $S1$, structured organization and cleaning are performed, including field unification and temporal alignment, removal of anomalous or low-quality samples, and normalization of tool-invocation logs. Sensitive content irrelevant to training is anonymized or removed. In addition, data can be further labeled by task phases, link states, and tool types. 

\textit{S3. Task Clustering: }
Unsupervised clustering is conducted on the curated prompts, action sequences, and tool-invocation chains to identify high-frequency and modularizable UAV task patterns. The key objective of this step is to automatically decompose generalized agent behaviors into a set of trainable, evaluable, and deployable aerial subtask clusters, providing clear targets for subsequent specialization of SLMs.

\textit{S4. SLM Selection: }
Candidate SLMs are selected based on on-board resource constraints and deterministic latency requirements, ensuring stable operation within limited computation and power budgets and maintaining controllable latency under workload fluctuations and link disturbances. On the UAV side, model families that are more amenable to quantization, and support faster decoding are preferred for fast decision-making. On the ground side, LLMs are retained for slow reasoning, global planning, and complex coordination. 

\textit{S5. LLM-to-SLM Distillation: }
On the ground side, LLMs serve as teachers to generate high-quality demonstrations and alignment signals for the typical task clusters identified in $S3$, including more accurate reasoning paths, more effective tool-selection strategies, and structured outputs. The goal of distillation is to enable SLMs to learn the decision distributions and key behavioral patterns of the LLM teachers, thereby increasing capability density without expanding parameter scale. 

\textit{S6. Continual Fine-tuning: }
Following distillation, the system continuously collects scenario data during real flights, including context summaries, tool-invocation trajectories, and task feedback under varying altitudes and speeds, link qualities, payloads, and unexpected events. These data are organized into specific training datasets. On this basis, the ground side periodically updates SLMs using parameter-efficient fine-tuning methods (e.g., LoRA or QLoRA), enabling the models to maintain robust outputs under environmental changes and task drift.

\subsection{Aerial Agentic AI Architecture}
\begin{figure*}[htbp]
	\centering
	\includegraphics[width=15cm]{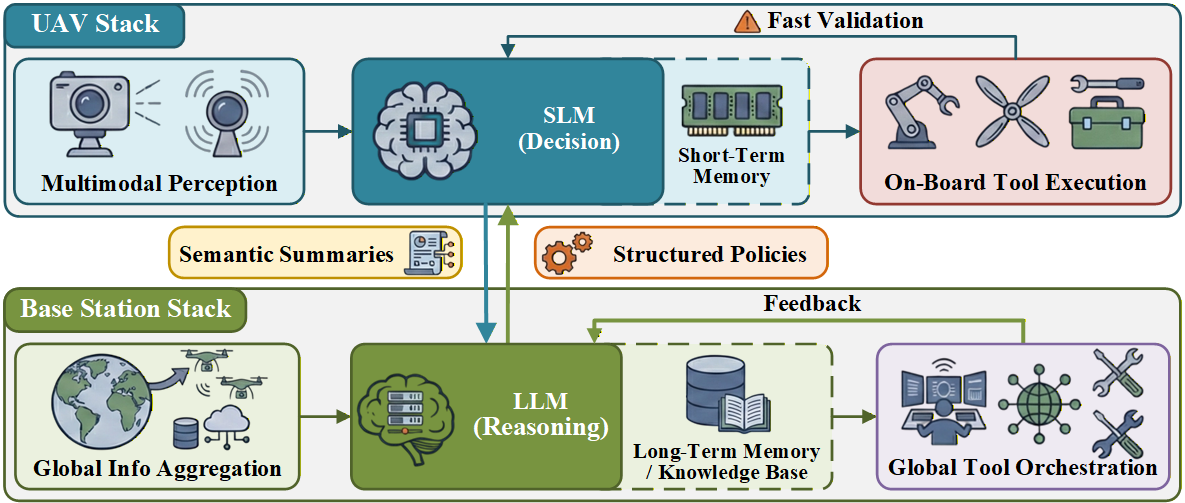}
	\caption{Aerial Agentic AI Architecture.}
	\label{fig:fig41}
\end{figure*}

In LAWNs, Aerial Agentic AI is best realized through a hierarchical paradigm of “fast closed-loop execution on the aerial side + slow reasoning enhancement on the ground side.” The specific structure of Aerial Agentic AI is as follows and is shown in Fig. \ref{fig:fig41}.

\subsubsection{UAV-Side Component Stack}\mbox{}\\
\hspace*{\parindent}\textit{a) Multimodal Perception:} Perception Module is responsible for preprocessing multi-source sensory inputs (e.g., cameras, radar, IMU, GNSS, and CSI) to form compact multimodal state representations for decision-making. This module emphasizes low latency and robustness, ensuring stable state inputs despite limited on-board computation and rapidly changing environments.

\textit{b) SLM-based Decision:} On-board inference is centered on SLMs, focusing on fast reasoning and executable outputs, such as trajectory adjustment, semantic summary generation, and target recognition and tracking. Rather than pursuing maximal general-purpose capability, the primary objective is to deliver stable and controllable decision outputs within strict latency budgets.

\textit{c) Short-Term Memory:} The UAV maintains short-term memory capturing recent observations, current intents, and local constraints to preserve task continuity and short-term decision consistency. This memory is typically implemented via lightweight caches, prioritizing fast read/write operations and supporting local closed-loop control and immediate interaction.

\textit{d) On-Board Tool Execution:} The UAV invokes controllable and deterministic on-board interfaces and lightweight operators, including sensor acquisition, flight-control APIs, lightweight detection and tracking operators, and local map caches. The core objective is to ensure interface compatibility, parameter normalization, and stable invocation order, enabling decisions to be promptly translated into executable actions.

\textit{e) Fast Validation:} Within strict constraints, the UAV performs rapid self-checking of final control commands, including logical consistency verification, boundary checking of critical parameters, and confidence screening. If anomalies or low-confidence outputs are detected, predefined fallback actions are executed to maintain a safe closed loop, while a collaboration request is issued to the BS. The ground side then performs further processing based on LLMs and long-term memory.

\subsubsection{BS-Side Component Stack}\mbox{}\\
\hspace*{\parindent}\textit{a) Global Information Aggregation:} The BS is responsible for receiving and aggregating semantic summaries from multiple UAVs. This module performs temporal alignment, conflict resolution, and multi-source fusion to produce a unified feature representation for global planning and coordinated scheduling. It provides stable and traceable inputs for subsequent LLM-based slow reasoning, tool-chain orchestration, and deep reflection.

\textit{b) LLM-based Reasoning:} The BS runs LLMs to perform cross–time-scale global planning and task decomposition, including global trajectory planning, cross-region coverage strategies, resource allocation strategies, and other tasks. Outputs are structured as task chains and executable instructions, and global state information is used to enforce consistency constraints across multi-UAV coordination strategies.

\textit{c) Long-Term Memory and Global Knowledge Base:} The BS maintains both long-term memory and a global knowledge base. Long-term memory focuses on experience accumulation, storing operational histories and strategy outcomes across time scales, supporting evaluation, reflection, and strategy optimization. The global knowledge base maintains general priors, organizing reusable structured knowledge to provide stable background information for planning, coordination, and tool orchestration. Together, these form the knowledge foundation underpinning BS-side LLMs.

\textit{d) Global Tool Orchestration:} The BS builds a global tool library and performs tool orchestration driven by slow reasoning for cross-task and cross-UAV coordination. The tool library includes high-computation, resource-intensive tools (e.g., code generation and auto-debugging, network-level simulation, algorithm evaluation and parameter tuning, and high-complexity optimizers). The LLM jointly reasons about task–tool–resource allocation from a global perspective, producing executable action sequences that UAVs can directly implement.

\textit{e) Reflection and Strategy Update:} the BS evaluates and reflects on critical feedback returned by UAVs, aggregating performance variations across different task phases and operating conditions to identify failure modes and strategy deviations. It then derives refined strategies (e.g., threshold/parameter updates, AI-weight adjustments, and short-term memory refresh) and encodes them in a structured, lightweight format, which is transmitted back to the UAV side for seamless updates. 

\section{\textcolor{black}{Case Study}}
\subsection{Experimental Setup}
\textcolor{black}{To validate the effectiveness of the Aerial Agentic AI architecture in LAWNs, we constructed a collaborative agent network simulation environment consisting of UAV aerial nodes and a BS ground node, and evaluated it on an urban monitoring and data backhaul task.} A heterogeneous deployment strategy was adopted for model selection: the UAV side deployed TinyLlama-1.1B \cite{zhang2024tinyllama} as a lightweight SLM to accommodate strict on-board computation and energy constraints, while the BS side deployed LLaMA2-7B \cite{touvron2023llama} as a slow-thinking LLM responsible for complex global reasoning and planning tasks.  The SLM side was equipped with lightweight path-planning operators (e.g., A*, D* Lite, etc.), while the LLM side used global planning algorithms (e.g., Genetic algorithm, PointerNet, etc.). Short-term memory was implemented using MemoryBuffer, and long-term memory/knowledge base was implemented using Milvus. The UAV hardware platform was a Jetson Orin NX 64GB, and the BS hardware platform consisted of an Intel Xeon Gold CPU (2.6 GHz) and 2 NVIDIA A100 GPUs (80 GB VRAM each).

First, we evaluated the impact of different optimization techniques on SLMs, including Activation-aware Weight Quantization (AWQ)\cite{lin2024awq}, OF \cite{salmani2025llm}, and FA \cite{dao2022flashattention}. Evaluation metrics included inference energy consumption and decoding speed on Jetson Orin NX. Comparison baselines included TinyLlama + AWQ, TinyLlama + OF, TinyLlama + FA, LLaMA2-7B + AWQ, and our method (TinyLlama + AWQ + OF + FA).

\begin{figure}[htbp]
	\centering
	\includegraphics[width=9cm]{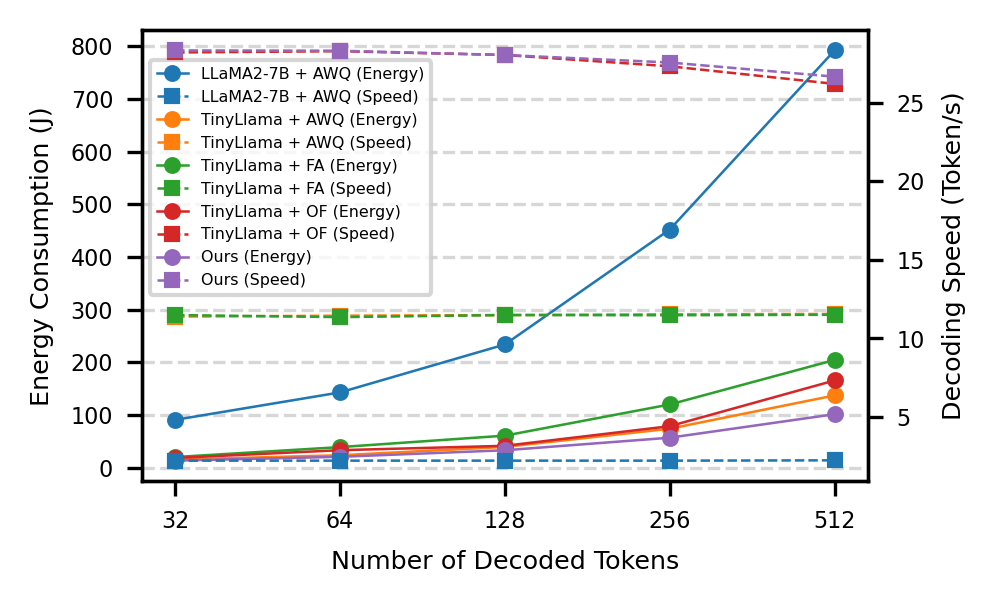}
	\caption{Energy Consumption and Decoding Speed of Different Methods.}
	\label{fig:fig5}
\end{figure}

Results in Fig. \ref{fig:fig5} indicate that even with AWQ quantization, LLaMA2-7B incurs inference energy consumption and decoding latency beyond the operational capacity of the Jetson Orin NX. In contrast, TinyLlama-1.1B is deployable on the UAV platform. Among the optimization techniques, AWQ significantly reduces inference energy, Fusion greatly accelerates decoding speed, and our combined approach balances both energy and latency. This demonstrates the feasibility of achieving a “high capability density” and “low-latency” trade-off for UAV-side deployment.
\begin{figure}[htbp]
	\centering
	\includegraphics[width=9cm]{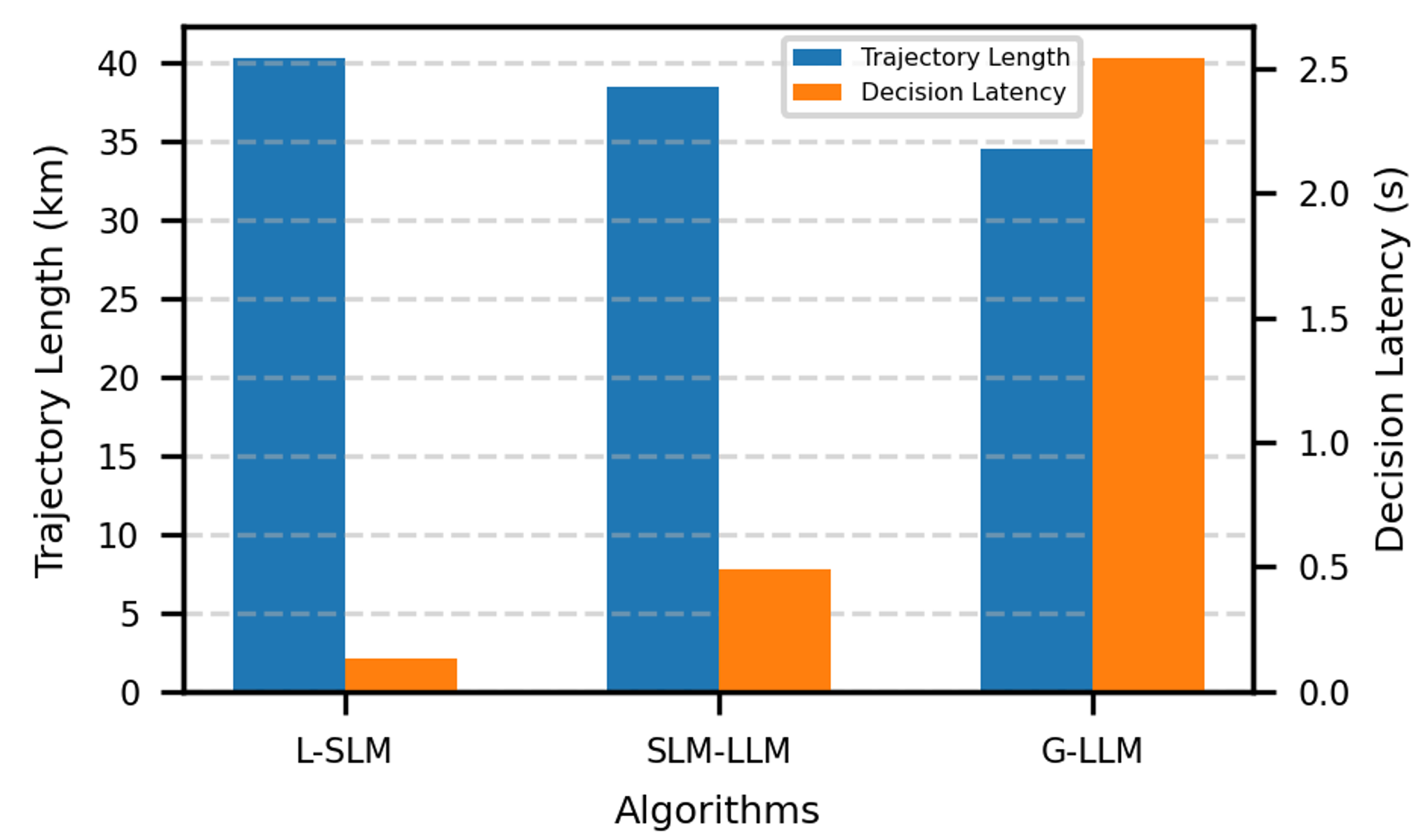}
	\caption{Trajectory Length and Decision Latency under Different Collaborative Modes.}
	\label{fig:fig6}
\end{figure}

Then, we evaluated the overall performance of Aerial Agentic AI. The simulation involved four UAVs and one BS, with UAV trajectory planning as the optimization task. Comparison strategies included local SLM decision-making (L-SLM), in which UAVs make decisions solely with TinyLlama on-board without BS assistance; global LLM decision-making (G-LLM), in which all data are sent to the BS and processed by LLaMA2; and collaborative SLM-LLM decision-making (SLM-LLM), in which UAVs make local decisions using TinyLlama and periodically synchronize with LLaMA2 at the BS. Evaluation metrics included decision latency and trajectory length.

Results in Fig. \ref{fig:fig6} indicate that local SLM decisions achieve the shortest latency but are prone to local optima in complex trajectory-planning tasks. LLM global decisions achieve the shortest total trajectory length but incur the highest latency. In the SLM-LLM collaborative scheme, UAVs perform nine independent SLM decisions before synchronizing with the BS, where long-term memory is updated and short-term memory and trajectory-planning strategies on the UAVs are adjusted based on global knowledge. This approach maintains low latency and near-optimal path quality even under unstable link conditions. 

\section{Open Issues}
\subsubsection{Collaboration Theory Gaps}
While fast local execution on UAVs and slow global reasoning at the BS is sensible in practice, a unified framework is still missing to determine when edge-side decision suffices and when BS-side enhancement is required. Future work should incorporate task risk, link conditions, energy/compute budgets, and safety constraints into a unified theoretical modeling framework to derive interpretable reasoning depth and adaptive aerial–ground collaboration policies, thereby moving collaboration beyond heuristic rules toward theoretically grounded solutions.

\subsubsection{Memory Drift Risks}
Combining UAV short-term memory with BS long-term memory may cause stale knowledge and error accumulation on UAVs under weak or intermittent links. Future research should advance semantic-level aerial–ground communication to enable fast memory synchronization and alignment, support trustworthy knowledge-slice updates, and strengthen privacy and anti-poisoning protections, keeping long-term evolution controllable and reliable.

\subsubsection{Tool Retrieval Limitations}
UAV-side SLMs face limited context and comprehension, leading to incomplete retrieval, tool confusion, or mismatches during invocation, which degrades decision accuracy. Future work should design efficient tool selection for aerial tasks: a lightweight index and capability tags can perform coarse Top-K filtering, then expose only a small candidate set under current resource constraints, improving tool-use accuracy and reliability.

\section{Conclusion}
This study presents an Aerial Agentic AI framework for LAWNs, built on fast-thinking UAV-side SLMs for execution and slow-thinking BS-side LLMs for reasoning. Through hierarchical collaboration, it combines real-time autonomy with global optimization: UAVs use lightweight models, short-term memory, and native API toolchains for rapid perception, local planning, and closed-loop control, while the BS leverages long-term memory and a global knowledge base for deep reflection, complex reasoning, knowledge updates, and high-computation tool orchestration. 
This work provides a technical foundation for future LAWNs to support self-organization, self-optimization, and self-evolution.

\bibliographystyle{IEEEtran}
\bibliography{ref}

\section*{Biographies}
\textbf{Li Dong} (Dlj2017@hunnu.edu.cn) is currently a Professor at Hunan University of Technology and Business, China.

\textbf{Feibo Jiang} (jiangfb@hunnu.edu.cn) is currently an Associate Professor at Hunan Normal University, China.

\textbf{Kezhi Wang} (Kezhi.Wang@brunel.ac.uk) is a Professor with the Department of Computer Science, Brunel University London, U.K.

\textbf{Cunhua Pan} (cpan@seu.edu.cn) is currently a full professor in Southeast University, China. 

\textbf{Dong In Kim} (dongin@skku.edu) is a Distinguished Chair Professor at Sungkyunkwan University, South Korea.

\textbf{Ekram Hossain} (Ekram.Hossain@umanitoba.ca) is currently a full professor in University of Manitoba, Canada. 
\end{document}